\documentclass[aps,prb,twocolumn]{revtex4}

\usepackage{amssymb}\usepackage{graphicx}\usepackage{hyperref}

\begin{document}

\title{Synthesis, Structure, and Ferromagnetism of a New Oxygen Defect
Pyrochlore System ${\rm \mathbf{Lu_2V_2O}}_{\mathbf{7-}x}$ ($x$ =
0.40--0.65)}

\author{G. T. Knoke}\altaffiliation[Present address: ]{Aret\'e Associates, 
P.O. Box 6024, Sherman Oaks, California 91413}

\author{A. Niazi}\altaffiliation[Permanent address: ]{Department of Physics, 
Faculty of Natural Sciences, Jamia Millia Islamia, New Delhi -- 110025, India}

\author{J. M. Hill}

\author{D.~C. Johnston}\affiliation{Ames Laboratory and Department of Physics
and Astronomy, Iowa State University, Ames, Iowa 50011}

\date{\today}

\begin{abstract}

A new fcc oxygen defect pyrochlore structure system ${\rm Lu_2V_2O}_{7-x}$
with $x$ = 0.40--0.65 was synthesized from the known fcc ferromagnetic
semiconductor pyrochlore compound $\rm{Lu_2V_2O_{7}}$ which can be written
as $\rm{Lu_2V_2O_{6}O^\prime}$ with two inequivalent oxygen sites O and
O$^\prime$.  Rietveld x-ray diffraction refinements showed significant Lu-V
antisite disorder for $x \gtrsim 0.5$.  The lattice parameter versus $x$
(including $x = 0$) shows a distinct maximum at $x \sim 0.4$.  We propose
that these observations can be explained if the oxygen defects are on the
O$^\prime$ sublattice of the structure.  The magnetic susceptibility versus
temperature exhibits  Curie-Weiss behavior above 150~K for all $x$, with a
Curie constant $C$ that \emph{increases} with $x$ as expected in an ionic
model.  However, the magnetization measurements also show that the
(ferromagnetic) Weiss temperature $\theta$ and the ferromagnetic ordering
temperature $T_{\rm C}$ both strongly \emph{decrease} with increasing $x$
instead of increasing as expected from $C(x)$.  The
$T_{\rm C}$ decreases from 73~K for $x = 0$ to 21 K for $x = 0.65$. 
Furthermore, the saturation moment at a field of 5.5~T at 5 K is nearly
independent of $x$, with the value expected for a fixed spin~1/2 per V\@. 
The latter three observations suggest that ${\rm Lu_2V_2O}_{7-x}$ may contain
localized spin~1/2 vanadium moments in a metallic background that is induced
by oxygen defect doping, instead of being a semiconductor as suggested by the
$C(x)$ dependence.

\end{abstract}


\maketitle

\section{\label{intro}Introduction}

The surprising discovery of heavy fermion
behaviors in the metallic fcc normal spinel structure $d$-electron compound
$\rm{LiV_2O_4}$ at low temperatures $T \lesssim10$\,K
(Refs.\ \onlinecite{Kondo1997, Johnston2000}) illustrates that highly
unconventional ground states can accrue to metallic compounds in which
geometric frustration for antiferromagnetic ordering is present within a
magnetic sublattice of the structure.  At high
$T \gtrsim 50$\,K the V spin $S =1/2$ sublattice interacts
antiferromagnetically and consists of corner-sharing tetrahedra with the
edges of the tetrahedra running along the six [110] crystal directions.  Each
V tetrahedron of course consists of triangles for which a collinear
antiferromagnetic ordering does not minimize the interaction energy of the
sublattice.  This geometric frustration has a tendency to suppress static
long-range antiferromagnetic ordering, and in the case of $\rm{LiV_2O_4}$
probably contributes to the formation at low temperatures of a heavy
fermion state instead. \\

\begin{figure}[t]
\includegraphics[width=3in]{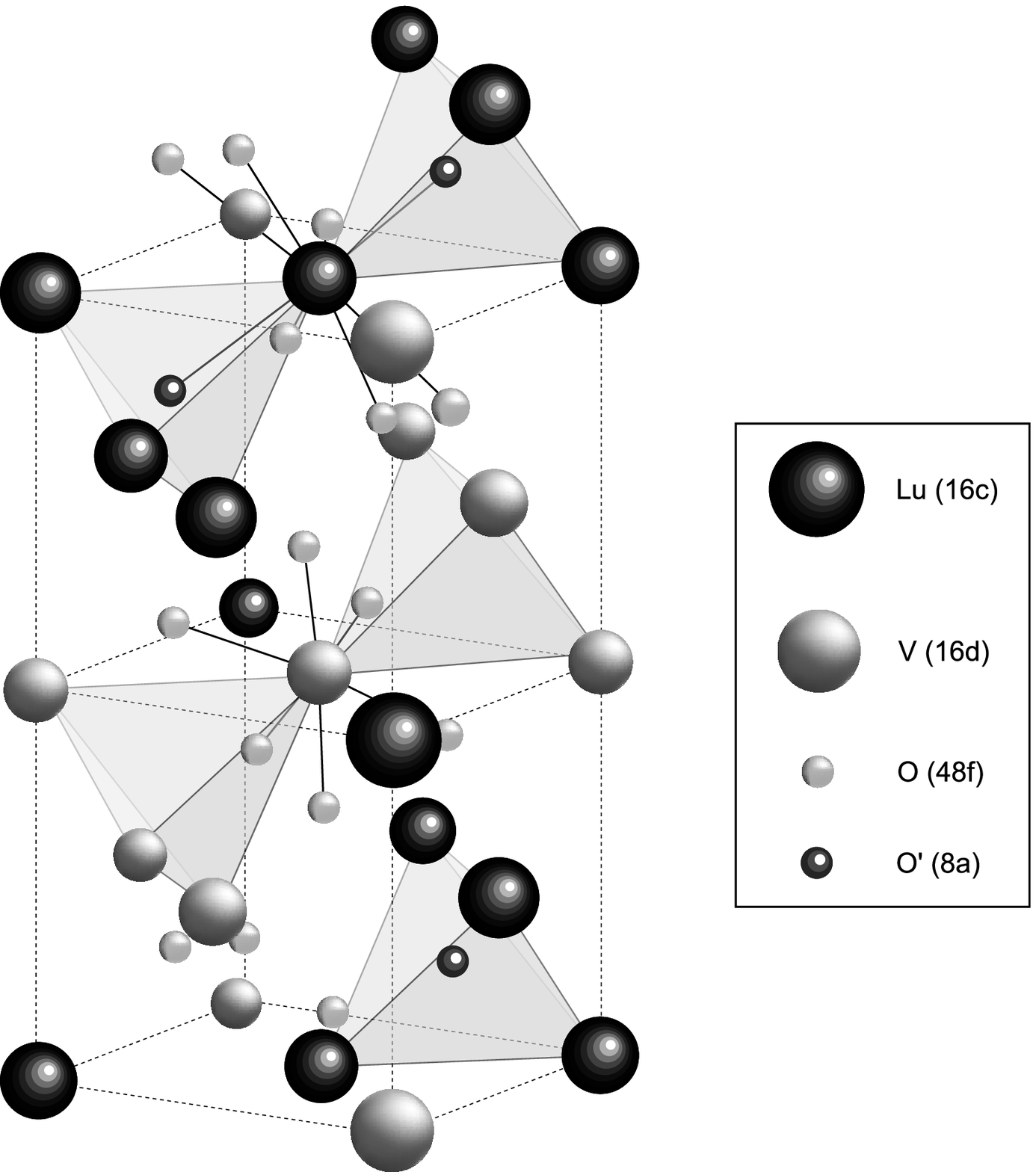}
\caption{Pyrochlore crystal structure of $\rm{Lu_2V_2O_7}$.}
\label{Fig-structure}
\end{figure}

In our search for the same or other novel ground states of such metallic
oxide systems containing frustrated arrays of local magnetic moments, we
turned to materials crystallizing in the well-known fcc pyrochlore structure
\cite{Subramanian1983} with generic formula
$\rm{A_2B_2O_7}$ in which the interpenetrating A and B sublattices are each
identical to the V sublattice in $\rm{LiV_2O_4}$.\cite{misnomer}  
The pyrochlore structure has been very popular and
important for the study of geometric magnetic frustration effects, but most 
such studies have been on insulating rather than metallic systems.  There are
two crystallographically inequivalent types of oxygen atoms O and O$^\prime$
in the structure, corresponding to the formula
$\rm{A_2B_2O_6O^\prime}$, as illustrated in Fig.~\ref{Fig-structure} for the
known ferromagnetic semiconducting pyrochlore compound
$\rm{Lu_2V_2O_7}$ with a Curie temperature $T_{\rm C}$ of about
73\,K\@.\cite{Kitayama1976,Bazuev1976,Bazuev1977,Shinike1977,Greedan1979,Soderholm1979,Soderholm1980,Soderholm1982}
This compound contains V$^{+4}$ $d^1$ cations with spin $S = 1/2$ which give
rise to the magnetic ordering (the Lu$^{+3}$ cations are nonmagnetic). 
Interestingly, the O$^\prime$ atoms are located at the centers of the
Lu$^{+3}$ tetrahedra.  The result of this is that the Lu site is eight-fold
coordinated by oxygen as compared to the six-fold coordination of V\@.  Thus
in $\rm{A_2B_2O_7}$ pyrochlore compounds the A atom is often a lanthanide
atom and the B atom is often a transition metal.  In the absence of the
O$^\prime$ atoms, both the Lu and V atoms would have been 6-fold coordinated
by oxygen; this is relevant to our proposed structural model for the oxygen
defects in ${\rm Lu_2V_2O}_{7-x}$ to be introduced later in
Sec.~\ref{SecOModel}.

In an attempt to dope the compound $\rm{Lu_2V_2O_7}$ to induce metallic
character, we tried to synthesize the unknown hypothetical compound
$\rm{Lu_2V_2O_6F}$ in which the O$^\prime$ atom in
$\rm{Lu_2V_2O_7}$ would be completely replaced by F, leading to a
crystallographically ordered compound with a non-integral oxidation state of
+3.5 for V as in $\rm{LiV_2O_4}$.  If the structure remained cubic, that
compound would have been metallic by symmetry.  Using conventional solid
state methods, a clean synthesis was carried out in a sealed Mo crucible at
1250 $^\circ$C, but unfortunately the product did not crystallize in the
pyrochlore structure. 

In a further attempt to induce a non-integral V oxidation state in
$\rm{Lu_2V_2O_7}$, we next tried to create oxygen vacancies that would
result in the composition ${\rm Lu_2V_2O}_{7-x}$; for $x = 1/2$, the oxidation
state of the V would again be +3.5 as in $\rm{LiV_2O_4}$.  These efforts were
successful as described herein, yielding compounds with $x = 0.40$--0.65. 
Our different types of magnetization data give conflicting
indications regarding the metallic character of ${\rm Lu_2V_2O}_{7-x}$, as
will be described.  In general, the  crystallographic and magnetic properties
evolve in interesting and unexpected ways with increasing~$x$.  Here we
report the synthesis, structure, and magnetic properties of the new
oxygen defect pyrochlore system ${\rm Lu_2V_2O}_{7-x}$.  The synthesis and
structural studies are presented in Sec.\ \ref{struct}, and the magnetization
measurements and analyses are in Sec.\ \ref{mag}.  A summary is given in
Sec.\ \ref{Summary}.

\section{\label{struct}Synthesis and Structure of
L\lowercase{u}$_{2}$V$_{2}$O$_{7-x}$ ($x$ = 0, 0.40 -- 0.65)} 

\subsection{Synthesis}

Lu$_{2}$V$_{2}$O$_{7}$ was synthesized by calcining pelletized stoichiometric
quantities of 99.995\% pure (metals basis) Lu$_{2}$O$_{3}$ (Stanford
Materials Corp.), V$_{2}$O$_{3}$ and V$_{2}$O$_{5}$ (MV Laboratories, Inc.)
in sealed silica tubes at 1250\,$^{\circ}$C for 72 hours with two
intermediate grindings. Samples of Lu$_{2}$V$_{2}$O$_{7-x}$ were
subsequently prepared by reducing powdered Lu$_{2}$V$_{2}$O$_{7}$ in a
mixture of 4.5\% H$_2$ in He at temperatures ranging from 550 to 
750$^{\circ}$\,C as listed in Table~\ref{tabStruct} below.  If the
reduction temperature was increased to 800~$^\circ$C, x-ray and magnetization 
data showed that the different compound LuVO$_3$ (see, e.g., Refs.\
\onlinecite{Nguyen1995,Munoz2004}) was obtained instead of pyrochlore
structure Lu$_{2}$V$_{2}$O$_{7-x}$.  The most effective method of synthesis
was to reduce a sample of Lu$_{2}$V$_{2}$O$_{7}$ under flowing H$_2$/He gas
in a Perkin-Elmer Thermogravimetric Analyzer (TGA).  Synthesis was regarded as
complete when weight loss ceased at a given temperature and the weight
remained stable for at least one hour.  The time to completion depended on a
combination of factors including the reduction temperature and the amount of
sample used.  Typical samples (40--50 mg) took approximately two days to
stabilize at a reduced composition, larger samples more than three days. 
The most homogeneous samples were produced by heating at a ramp rate of
2\,$^{\circ}$C/min followed by a two-day hold and a subsequent
2\,$^{\circ}$C/min cooling rate with a 15~min hold after room temperature had
been reached to allow the TGA to stabilize.  Two series of 
Lu$_{2}$V$_{2}$O$_{7-x}$ samples, ``gtk-4-11-n'' and ``gtk-4-5c2-n'', were
synthesized and studied. 

The oxygen content of the parent Lu$_{2}$V$_{2}$O$_{7}$ samples was
determined via oxidation to V$^{+5}$ in the TGA under O$_2$.  A typical value
of the oxygen content was 6.98(7).  For the oxygen deficient
Lu$_{2}$V$_{2}$O$_{7-x}$ samples, the weight loss during synthesis in the TGA
was used to calculate the oxygen content, assuming the parent compound had
exactly $x = 0$. Attempts to produce Lu$_{2}$V$_{2}$O$_{7-x}$ in a tube
furnace under flowing H$_2$ in He were largely unsuccessful.  The precise
temperature of the sample within the furnace was difficult to measure and it
was impossible to know when the reduction had ceased without a constant
monitor on the weight. Checking the weight periodically by removing the
sample from the furnace proved disastrous as repeated heating and cooling
cycles caused the samples to decompose.  These samples had multiple Curie
temperatures, were generally of poor quality, and will not be further
discussed.    In view of these results and the relatively low temperatures
(550--750~$^\circ$C) at which the samples are synthesized compared to the
temperature (1250~$^\circ$C) needed to synthesize the Lu$_{2}$V$_{2}$O$_{7}$
parent compound, the oxygen defect system Lu$_{2}$V$_{2}$O$_{7-x}$ may be
metastable.

\subsection{Structure}

\begin{figure}[t]
\includegraphics[width=3in]{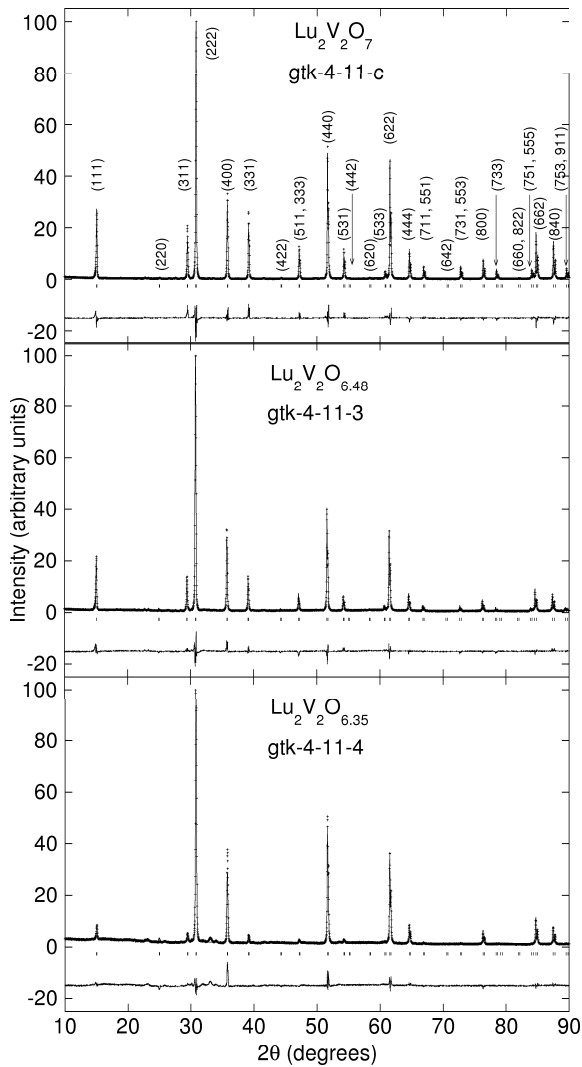}
\caption{Indexed X-ray powder diffraction patterns obtained using CuK$\alpha$
radiation, their Rietveld refinement fits, the expected line positions
(ticks), and the differences between observed and calculated 
intensities, for three single-phase samples of Lu$_2$V$_2$O$_{7-x}$ with $x$
= 0.00, 0.52, and 0.65. The strong decrease in the (odd odd odd) reflection
intensities with increasing $x$ results from Lu-V antisite disorder.  The
goodnesses of fit were $R_{\rm wp}=15.0$\% for $x = 0$, $R_{\rm wp} = 12.3$\%
for $x = 0.52$, and $R_{\rm wp} = 11.0$\% for $x = 0.65$.  The respective
$R_{\rm wp}/R_{\rm p}$ values are given in Table~\protect\ref{tabStruct}.}
\label{FigRietveld}
\end{figure}

Powder x-ray diffraction (XRD) patterns were obtained using a Rigaku
Geigerflex diffractometer and Cu K$\alpha$ radiation, in the 2$\theta$ range
from 5 to 110$^{\circ}$ with a 0.02$^{\circ}$ step size.  Intensity data were
accumulated for 5 to 10~s per step.  The small powder sample amounts
available ($\lesssim 50$~mg, see above) constrained us to mount the powder on
vaseline-coated glass microscope slides for XRD specimens.  The amorphous
vaseline and glass background at low angles restricted subsequent Rietveld
refinement of some of the XRD patterns to $2\theta > 20^{\circ}$.  However,
the restricted range eliminated only the initial (111) reflection and did not
significantly affect the results.  A typical sample of
Lu$_{2}$V$_{2}$O$_{7-x}$ showed a single phase x-ray diffraction pattern with
lattice parameters similar to those of the parent compound as shown for
several samples in Fig.\ \ref{FigRietveld}.  It is important to note that the
intensities of the reflections with Miller indices $(h~k~l)$ all odd become
greatly suppressed with increasing oxygen deficiency above $x\approx 0.4$
compared to the reflections with $(h~k~l)$ all even.  This reduction in the
(odd, odd, odd) reflection intensities is a  very sensitive indicator of
antisite disorder between the Lu and V sites of the pyrochlore structure as
reported for (Sc,Lu)--V antisite disorder in the system
(Lu$_{1-x}$Sc$_x){\rm _2V_2O_7}$.\cite{Greedan1979}  This means that Lu and V
atoms switch places to increasing degrees with increasing oxygen
deficiency~$x$.

\begin{table}

\caption{\label{tab-xrd-data1}
Crystal data for Lu$_{2}$V$_{2}$O$_{7-x}$}
\begin{ruledtabular}
\begin{tabular}{ll}
Space Group & {\it Fd$\overline{\it{3}}$m} (\#227), origin at center
($\overline{\it{3}}$\it{m})\\
$Z$ & 8 	\\
Atomic positions	& Lu, 16(c), ($\overline{\it{3}}$\it{m}), \rm{(0,0,0)} \\
& V, 16(d), ($\overline{\it{3}}$\it{m}),
(${1}\over{2}$,${1}\over{2}$,${1}\over{2}$) \\
& O, 48(f), \emph{mm}, (${1}\over{8}$, ${1}\over{8}$, $z$) \\
& O$^{\prime}$, 8(a), \emph{43m}, (${1}\over{8}$, ${1}\over{8}$,
${1}\over{8}$)\\
Wavelength & 1.54060~\AA \\
$2\theta$ range & $5 - 110^{\circ}$  \\
Number of reflections & 90 \\
Profile function & pseudo-Voigt \\
\end{tabular}
\end{ruledtabular}

\end{table}

\begin{table*}

\caption{\label{tabStruct}
Structure parameters for Lu$_{2}$V$_{2}$O$_{7-x}$ refined from powder
XRD data.  $T_{\rm prep}$ is the preparation temperature.  The overall
isotropic thermal parameter $B$ is defined within the temperature
factor of the intensity as $e^{-2B \sin^2 \theta/ \lambda^2}$.} 
\begin{ruledtabular}
\begin{tabular}{lrlcclccc}
Sample & $T_{\rm prep}$ & $x$ & a (\AA)& $z$ of & Lu-V &
V-O-V  & $B$  & $R_{\rm wp}/R_{\rm p}$\\
gtk-4- & ($^\circ$ C) & & (\AA) & O at 48(f) & Disorder & ($^{\circ}$) &
(\AA$^2$)\\\hline  
5-c2 &  1250 & 0 & 9.9368(1) & 0.4258(5) & 0\footnotemark[1] &	134.5 & 0.62(2)
& 1.30\\
5-c2-7\footnotemark[2] & 550 & 0.44 & 9.9721(3) & 0.4155(9) & 0.003(2)
&128.9 & 2.47(4) & 1.33\\  
5-c2-5\footnotemark[2] & 650 & 0.52 & 9.9643(2) & 0.4090(8) &
0.052(2) &125.5 & 1.92(3) & 1.32\\
5-c2-6 & 700 & 0.65 & 9.9502(3) & 0.4067(6) & 0.254(2) & 124.4 & 2.79(3) &
1.36\\
11-c &	1250 & 0 & 9.9401(1)& 0.4265(5) & 0\footnotemark[1] & 134.8  & 1.03(2)
& 1.32\\
11-1\footnotemark[2] & 550 & 0.40 &	9.9618(2) &	0.4190(9) & 0.000(2)
&130.6 & 2.94(4) & 1.31\\  
11-2\footnotemark[2] & 600 & 0.48 &	9.9604(3) &	0.4141(8) &
0.026(2) &128.1 & 3.12(4) & 1.37\\
11-3 & 650 & 0.52 & 9.9560(2) &	0.4114(6) & 0.085(2) & 126.7 & 3.12(3) &
1.33\\
11-5 & 700 & 0.58 & 9.9526(3) & 0.4057(7) & 0.145(2) & 123.8 & 2.73(3) &
1.32\\
11-4 & 750 & 0.65 & 9.9424(3) & 0.3912(8) & 0.295(2)& 116.9 & 2.72(3) &
1.28\\
\end{tabular}
\end{ruledtabular}

\footnotetext[1]{Rietveld refinement of the XRD data gave small unphysical
negative values ($\approx -0.01$) for the Lu-V antisite disorder, i.e., for 
site occupancy by the minority species [V at the Lu 16(c) site and Lu at the V
16(d) site].  The antisite disorder was therefore fixed to be zero for
subsequent refinements.}

\footnotetext[2]{Two-phase sample with about 20 mol\% or less (from Reitveld
refinement of the XRD data) of stoichiometric Lu$_{2}$V$_{2}$O$_{7}$. The
overall composition $x$ and the refined fraction of the
Lu$_{2}$V$_{2}$O$_{7-x}$ phase in each two-phase sample were 0.35 and 79(2)\%
for gtk-4-5-c2-7, 0.49 and 95(1)\% for gtk-4-5-c2-5, 0.32 and 80(1)\% for
gtk-4-11-1, and 0.39 and 82(1)\% for gtk-4-11-2.  The composition and
structure parameters reported in the table are for the oxygen defect
Lu$_{2}$V$_{2}$O$_{7-x}$ phase in each sample.}

\end{table*}

To quantitatively characterize the changes in the structure of
Lu$_{2}$V$_{2}$O$_{7-x}$ with increasing~$x$, Reitveld refinements of the
x-ray diffraction data were carried out using the program
DBWS9807a.\cite{dbws} The atomic positions of the atoms
in the pyrochlore structure of Lu$_{2}$V$_{2}$O$_{7}$ are shown in
Table~\ref{tab-xrd-data1}, together with several parameters associated with
the Rietveld refinements.  Table~\ref{tab-xrd-data1} shows that the Lu, V,
and O$^\prime$ positions are all fixed with respect to the unit cell edges,
and only the $z$ coordinate of the O position is variable. The initial O
$z$-coordinate was taken to be that reported for Lu$_{2}$V$_{2}$O$_7$ by
Soderholm and Greedan.\cite{Soderholm1982}  Keeping in mind the limitations
of our XRD data, the TGA results were used to determine the oxygen site
occupancy, with the vacancies being ascribed to the O$^{\prime}$ site (see
the structural model in Sec.~\ref{SecOModel} below).  Thus we did not refine
the oxygen site occupancies.  Similarly, the quality of the XRD patterns was
such that only an overall isotropic thermal parameter $B$ was refined. 
Furthermore, the surface roughness was not refined, which can affect the
fitted values of
$B$.  The parameters refined were $B$, scale, sample height, sample
transparency, background, lattice spacing, the O position $z$ value,
pseudo-Voigt profile shape, full-width at half-maximum, and the Lu-V antisite
disorder.  The XRD patterns of samples with overall $x < 0.5$ showed the
presence of stoichiometric  Lu$_{2}$V$_{2}$O$_{7}$ ($\sim$ 20 mol\% or less)
in addition to Lu$_{2}$V$_{2}$O$_{7-x}$, probably due to kinetic limitations
during the reduction process, and were refined as 2-phase mixtures.  The
reported structure parameters for Lu$_{2}$V$_{2}$O$_{7-x}$ are those refined
for this phase.  Bond lengths and angles were computed from the refined
structure parameters using PowderCell for Windows.\cite{pcell}  The XRD
refinement parameters are shown in Table~\ref{tabStruct}. 
The Reitveld refinement fits for single phase samples with $x = 0$, 0.52 and
0.65 are shown in Fig.~\ref{FigRietveld}.   

\begin{figure}[t]
\includegraphics[width=3.3in]{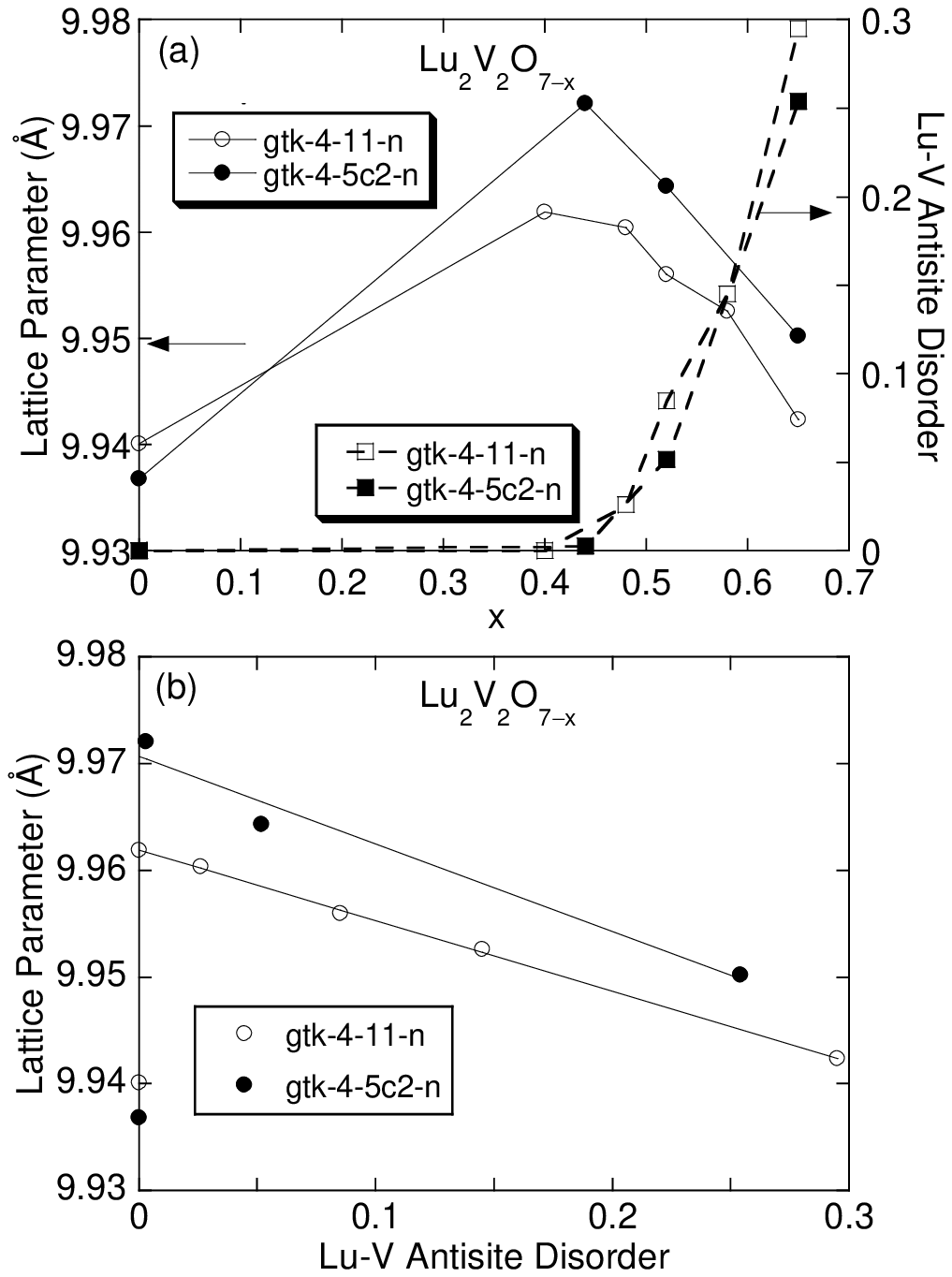}
\caption{(a) Refined lattice parameter (left scale) and Lu-V antisite
disorder (right scale) as functions of $x$ for two series of
Lu$_{2}$V$_{2}$O$_{7-x}$samples. (b) Refined lattice parameter as a function
of the Lu-V antisite disorder. The straight lines are linear fits to the data
for $x > 0.4$.  The error bars are smaller than the size of the
data symbols.} 
\label{xrd-parameters}
\end{figure}

The lattice parameter and Lu-V antisite disorder are plotted as functions of
the oxygen deficiency $x$ for the two series of Lu$_{2}$V$_{2}$O$_{7-x}$
samples in Fig.~\ref{xrd-parameters}(a). The lattice parameter varies
nonmonotically, initially increasing and then decreasing for increasing $x$,
thus strongly violating Vegard's law.  In the absence  of samples with $0 < x
< 0.4$ it is not possible to pinpoint the $x$ value  corresponding to the
maximum in the lattice parameter, but similar behavior was observed for both
series of samples.  The refined Lu-V  antisite disorder in
Fig.~\ref{xrd-parameters}(a) becomes significant for $x \gtrsim 0.5$ and then
increases monotonically with increasing $x$.  The lattice parameter decreases
approximately linearly with the Lu-V antisite disorder for $x\geq 0.4$ as
shown in Fig.~\ref{xrd-parameters}(b).  

\subsection{Structural Model for Oxygen Vacancies in ${\rm
\mathbf{Lu_2V_2O}}_{\mathbf{7-}x}$\label{SecOModel} }

Typically, in oxides V$^{+3}$ or V$^{+4}$ cations are 6-fold coordinated
by oxygen whereas the larger Lu$^{+3}$ cation is often 8-fold coordinated, as
in the pyrochlore structure of Lu$_{2}$V$_{2}$O$_{7}$ itself, although 6-fold
coordination for Lu also occurs in some oxide compounds.  In view of the Lu-V
antisite disorder in Lu$_{2}$V$_{2}$O$_{7-x}$ that increases with increasing
$x$, and of the nonmonotonic variation in the lattice parameter with $x$, we
propose that the oxygen vacancies in ${\rm Lu_2V_2O}_{7-x}$ occur on the
O$^\prime$ sublattice of the structure, as follows. Since the O$^\prime$
anions reside at the centers of the Lu tetrahedra in Lu$_{2}$V$_{2}$O$_{7}$
as shown in Fig.~\ref{Fig-structure}, O$^\prime$  vacancies would decrease
the average coordination number of the Lu sites by O, thus encouraging V atoms
to occupy the Lu sites, which requires Lu to switch places with V (antisite
disorder) in order to preserve the overall composition.  If both of the
O$^\prime$ atoms on either side of a given Lu site were vacant, which would
occur with probability $x^2$, that Lu site would then have 6-fold oxygen
coordination as preferred by V cations.  The antisite disorder in
Fig.~\ref{xrd-parameters} is indeed seen to roughly follow an
$x^2$ dependence.  

In this model, we propose that the removal of oxygen from the O$^\prime$ sites
at the centers  of the Lu tetrahedra causes an initial increase in the lattice
parameter with increasing $x$ from $x = 0$ to $x = 0.32$ in
Fig.~\ref{xrd-parameters}(a).  This arises from the resultant removal of the
Lu-O$^\prime$-Lu bonding within the affected Lu$_4$O$^\prime$ tetrahedra,
thus causing the Lu$_4$ tetrahedra and the overall lattice to expand. 
However, at larger $x \gtrsim 0.4$, the smaller ionic radius of the Lu$^{+3}$
ion in the 6-fold oxygen-coordinated V site, occurring via Lu-V antisite
disorder, compared to its ionic radius in the original 8-fold coordinated Lu
site, evidently leads to the observed lattice contraction.  For
$x = 0.40$--0.65, the lattice parameter decreases approximately linearly by
about 0.02 \AA\ with increasing Lu-V antisite disorder as seen in
Fig.~\ref{xrd-parameters}(b).  Indeed, this decrease is about the same as
expected from the decrease in ionic radius of Lu$^{+3}$ in 8-fold (0.977
\AA) to 6-fold (0.861 \AA) coordination by oxygen with increasing Lu-V
antisite disorder, where one utilizes the facts that there are 16 Lu cations
per unit cell and that the fractional Lu occupation of V sites increases from
about 0 to 0.3 over this $x$ range.

\section{\label{mag}Magnetization Measurements and Analyses}

\subsection*{Measurement Results}

\begin{figure}[t]
\includegraphics[width=3.3in]{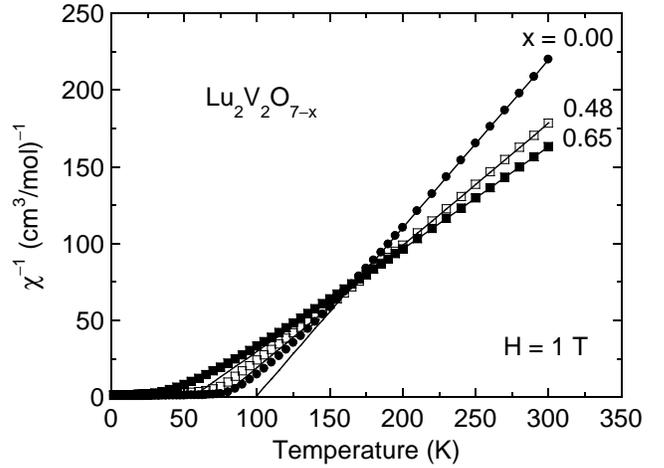}
\caption{Inverse of the magnetic susceptibility $\chi$ versus
temperature in a magnetic field of 1\,T for illustrative Lu$_2$V$_2$O$_{7-x}$
samples with $x = 0$, 0.48, and 0.65.  The Curie-Weiss law fits above 150\,K
are shown as straight lines and are extrapolated to lower temperatures.  The
sample with (corrected) $x = 0.48$ contains 20(1)\% of unreduced
Lu$_2$V$_2$O$_{7}$ which has been corrected for in the plotted data.} 
\label{chi-inv}
\end{figure}

\begin{table}

\caption{\label{tab-magdata}
Magnetic properties of Lu$_{2}$V$_{2}$O$_{7-x}$. Here, $T_{\rm C}$ is
the Curie temperature; $C$ the Curie constant and $\theta$ the Weiss
temperature in the Curie-Weiss behavior of the high-temperature
susceptibility; and $\mu_{\rm sat}$(5~K) the saturation moment at 5\,K.}
\begin{ruledtabular}
\begin{tabular}{cccccc}
Sample & $x$ & $T_{\rm C}$ & $C$ & $\theta$ &
$\mu_{\rm sat}$(5\,K)\\ 
(gtk-)& & (K)  & ${\rm (cm^3\,K/mol)}$ & (K) & $(\mu_{\rm B}$/V
atom) \\ \hline
4-5-c2 & 0.00 & 73.8(1) & 0.923(2) & 97.4(3) & 0.97\\
4-5-c2-7\footnotemark[1] & 0.44 & 39.0(4)\footnotemark[1] &
1.31(2)\footnotemark[1] & 72.0(3)\footnotemark[1] & 0.95\\
4-5-c2-5\footnotemark[1] & 0.52 & 35.2(6)\footnotemark[1]
&\footnotemark[2] &\footnotemark[2] &\footnotemark[3]\\
4-5-c2-6 & 0.65 &
20.8(2) &\footnotemark[2] &\footnotemark[2] &\footnotemark[3]\\
4-11-c & 0.00 & 72.0(1) & 0.825(1) & 97.3(2) & 0.95\\
4-11-1\footnotemark[1] & 0.40 & 37.1(4)\footnotemark[1] &
1.20(1)\footnotemark[1] & 79.8(5)\footnotemark[1] & 0.99\\
4-11-2\footnotemark[1] & 0.48 & 36.1(12)\footnotemark[1] &
1.37(1)\footnotemark[1] & 71.5(5)\footnotemark[1] & 0.93\\  
4-11-3 & 0.52 & 32.2(1) &
1.396(1) & 66.9(2) & \footnotemark[3]\\  
4-11-5 & 0.58 & 27.9(1) & 1.457(1) & 62.6(1) & \footnotemark[3]\\
4-11-4 & 0.65 & 20.3(2) & 1.500(1) & 55.4(1) & \footnotemark[3]\\

\footnotetext[1]{This sample shows a second ferromagnetic
transition at 73\,K indicating a significant fraction of the sample is
Lu$_{2}$V$_{2}$O$_{7}$ (see Table~\protect\ref{tabStruct}).  The listed
$x$ is the corrected oxygen defect composition of the Lu$_{2}$V$_{2}$O$_{7-x}$
phase.  Since the saturation moment does not significantly depend on the
oxygen vacancy composition $x$ in Lu$_{2}$V$_{2}$O$_{7-x}$, the mixture of
Lu$_{2}$V$_{2}$O$_{7}$ with Lu$_{2}$V$_{2}$O$_{7-x}$ does not affect the
value of the saturation moment.  The observed $C$ and $\theta$ values have
been corrected so that the listed values correspond to the
Lu$_{2}$V$_{2}$O$_{7-x}$ phase by itself.  This was done by correcting the
observed susceptibility for that of the relevant amount of the $x=0$ phase and
fitting the corrected data above 150~K by a Curie-Weiss law.}
\footnotetext[2]{The magnetic susceptibility at $H = 1$~T was not measured for
this sample.}
\footnotetext[3]{The magnetization of this sample does not completely saturate
up to an applied magnetic field $H = 5.5$\,T.}

\end{tabular}
\end{ruledtabular}

\end{table}

Magnetization $M$ data were
obtained at temperatures $T$ between 1.8 and 300 K over the applied
magnetic field $H$ range from 0 to 5.5~T using a Quantum Design SQUID
magnetometer.  As shown for representative samples in Fig.~\ref{chi-inv}, the
inverse susceptibility of each Lu$_{2}$V$_{2}$O$_{7-x}$ composition at
temperatures above about 150 K follows the Curie-Weiss law $\chi = C/(T -
\theta)$ with Curie constant $C$ and Weiss temperature $\theta$ listed in
Table~\ref{tab-magdata}.  Representative straight-line fits are shown in
Fig.~\ref{chi-inv}.  

\begin{figure}[t]
\includegraphics[width=3in]{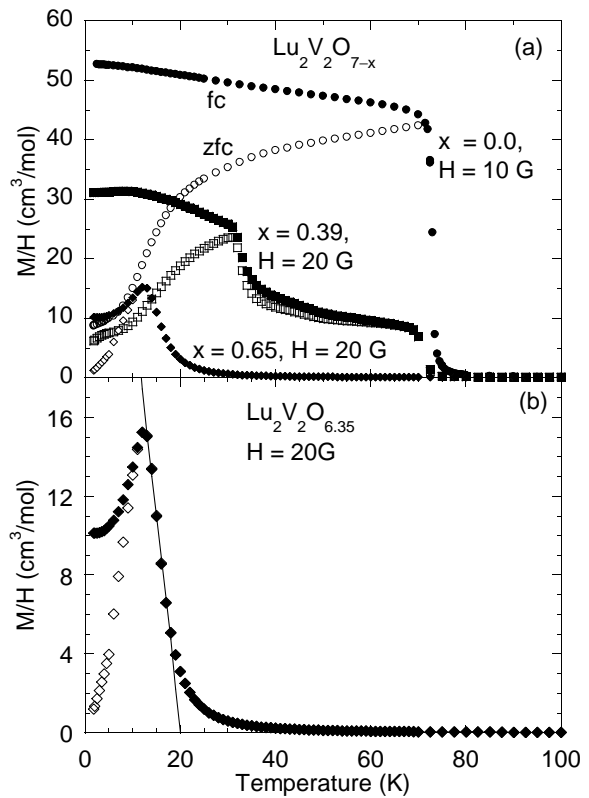}
\caption{(a) Zero-field-cooled
(zfc, open symbols) and field-cooled (fc, filled symbols) magnetization
versus temperature $M(T)$ data in low magnetic fields $H$ for $x= 0.0$, 0.39,
and 0.65.  The sample with overall $x = 0.39$ contained $18(2)$\% of
unreduced $\rm{Lu_2V_2O_{7}}$, as seen in the corresponding $T_{\rm C}
\approx 73$~K\@.  (b) Expanded plots of the fc and zfc data for the $x =
0.65$ sample, showing the straight-line fit to the $M(T)$ data to determine
the ferromagnetic ordering temperature $T_{\rm C}$.} 
\label{lowfdM}
\end{figure}

\begin{figure}[t]
\includegraphics[width=3in]{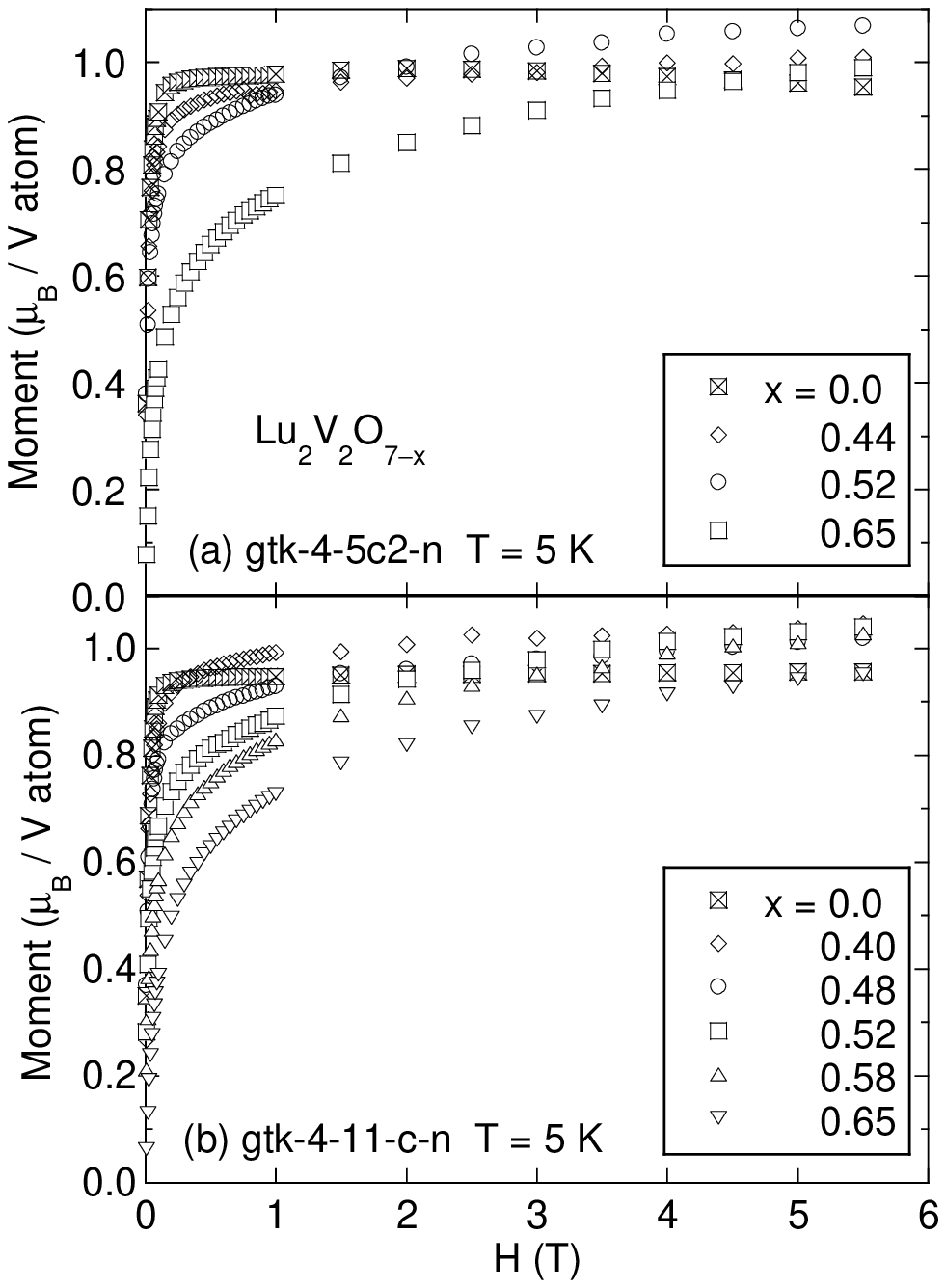}
\caption{Magnetization versus field at 5\,K for the two series of
Lu$_{2}$V$_{2}$O$_{7-x}$ samples (a) gtk-4-5-c2-n and (b) gtk-4-11-n. The
samples with $x \gtrsim 0.5$ did not completely saturate up to at least
5.5\,T\@.  The samples with listed compositions $x = 0.44$ and 0.52 in (a),
and $x = 0.40$ and 0.48 in (b), contain unreduced Lu$_2$V$_2$O$_{7}$ for which
the listed compositions are the actual corrected $x$ values in the oxygen
defect phase Lu$_{2}$V$_{2}$O$_{7-x}$.} 
\label{M-sat}
\end{figure}

All ${\rm Lu_2V_2O}_{7-x}$ samples ($x = 0.40$--0.65) exhibited ferromagnetic
ordering.  The Curie temperatures $T_{\rm C}$ (Table~\ref{tab-magdata}) were
determined from zero-field-cooled and field-cooled $M(T)$ measurements at a
low field $H = 10$\,G or 20\,G, shown in Fig~\ref{lowfdM}(a) for a few of the
samples.  The straight line fit to the $M(T)$ data used to determine $T_{\rm
C}$ is exemplified for the $x = 0.65$ sample in Fig~\ref{lowfdM}(b). 
Isothermal magnetization $M(H)$ curves at $T=5\,K$ are plotted in
Fig.~\ref{M-sat}.  The samples with $x \gtrsim  0.5$, containing significant
Lu-V antisite disorder, did not completely saturate up to at least $H=5.5$
T\@.  The well-defined saturation moments $\mu_{\rm sat}$ for samples with $x
< 0.5$ are listed in Table~\ref{tab-magdata}.  Assuming a spectroscopic
splitting factor (``g-factor'') $g = 2$, the saturation moment of the $x=0$
sample is close to the expected value $\mu_{\rm sat} = gS\mu_{\rm B} =
1\,\mu_{\rm B}$/V atom, since $S = 1/2$ for the $V^{+4}$ cations in that
compound.  On the other hand, the magnetizatons of all the other samples are
also clustered around 1 $\mu_{\rm B}$/V atom at our highest field of 5.5~T,
contrary to expectation (below) that $\mu_{\rm sat}$ should increase with
increasing $x$.

\begin{figure}[t]
\includegraphics[width=3.3in]{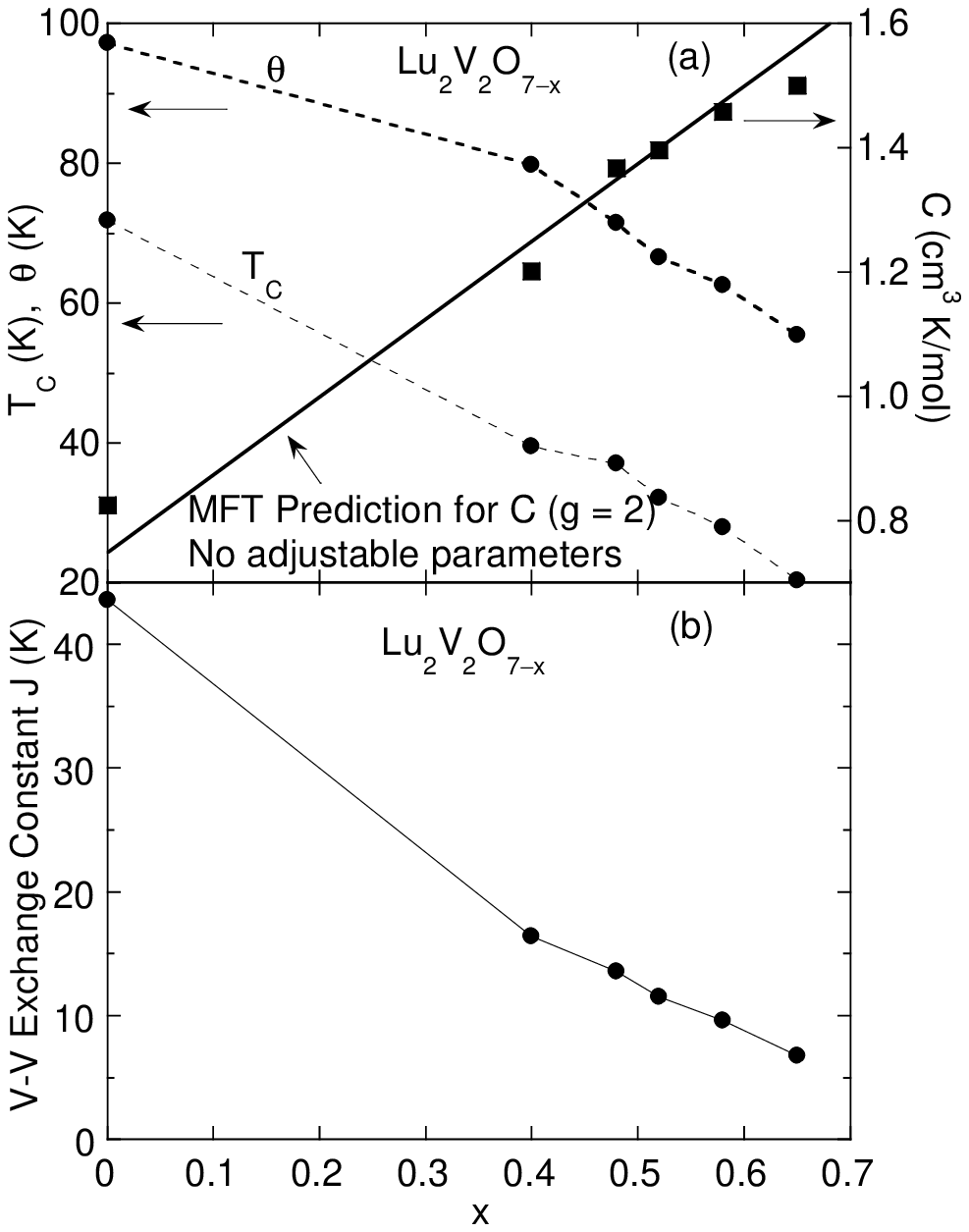}
\caption{(a) Curie temperature $T_{\rm C}$, Weiss temperature $\theta$ (left
scale) and Curie constant $C$ (right scale) versus oxygen deficiency $x$
in $\rm{Lu_2V_2O_{7-x}}$ series gtk-4-11-n.  The straight line is the mean 
field theory (MFT) prediction for $C$ in Eqs.\ (\protect\ref{eqC(x)})
and (\protect\ref{eqC(0)}) assuming a spectroscopic splitting factor $g = 2$
for the local magnetic moments in the system, with no (other) adjustable
parameters.  (b) Variation of the exchange constant $J$ versus composition $x$
in $\rm{Lu_2V_2O_{7-x}}$, computed using Eq.\ (\protect\ref{eqJ2}) and the $C$
and $T_{\rm C}$ data in Table~\protect\ref{tab-magdata}.} 
\label{Tc-theta-C}
\end{figure}

The $T_{\rm C}$, $\theta$ and $C$ are plotted in Fig.~\ref{Tc-theta-C}(a)
versus oxygen deficiency $x$ in ${\rm Lu_2V_2O}_{7-x}$, where for samples
with $x < 0.5$ containing unreduced ${\rm Lu_2V_2O}_{7}$, the Curie
constants and Weiss temperatures have been corrected to reflect the values of
the ${\rm Lu_2V_2O}_{7-x}$ phase in those three samples.  One sees that as
the Curie constant $C$ increases, both the Weiss temperature
$\theta$ and the $T_{\rm C}$ decrease, contrary to expectation (see below). 
The $\theta$ for each sample is consistently $\sim 30$~K higher than
$T_{\rm C}$, suggesting that $T_{\rm C}$ is suppressed from the mean-field
value $T_{\rm C} = \theta$ by fluctuation effects.

\subsection*{Magnetization Data Analysis}

In the Curie-Weiss law $\chi = C/(T - \theta)$, the Curie constant $C$ of a
system of $N$ spins $S$ is expressed as 
\begin{equation}
C = \frac{N g^2 S(S+1) \mu_{\rm B}^2}{3 k_{\rm B}}\, ,
\label{eqCurieConst}
\end{equation}
where $\mu_{\rm B}$ is the Bohr magneton and $k_{\rm B}$ is Boltzmann's
constant.  The Curie-Weiss law is a mean-field expression arising from only
nearest-neighbor magnetic interactions and correlations between the spins in
the system.  For a system containing a random distribution of different spin
values $S$, in mean field approximation one
can replace $S(S+1)$ in Eq.\ (\ref{eqCurieConst}) by its average value
$\langle S(S+1)\rangle$, i.e. 
\begin{equation}
C = \frac{N g^2 \langle S(S+1)\rangle \mu_{\rm B}^2}{3 k_{\rm B}}\, .
\label{eqCurieConst2}
\end{equation}

In ${\rm Lu_2V_2O}_{7-x}$, the formal oxidation state of the V ions is $4 -
x$. If the reduction in oxidation state with $x$ does not give rise to
itinerant electrons but rather to a random mixture of localized $S = 1/2$
(V$^{+4}$) and $S = 1$ (V$^{+3}$) spins, then in this ionic model the
probabilities that $S = 1/2$ and $S = 1$ occur are $p_{1/2} = 1 - x$ and $p_1
= x$, respectively, yielding  
\begin{equation}
\langle S(S+1)\rangle = \frac{3 + 5 x}{4}\, .
\label{eqSS1}
\end{equation}
If the $g$-factors of the spins 1/2 and 1 are the same, one
can substitute Eq.~(\ref{eqSS1}) into Eq.~(\ref{eqCurieConst2}) to obtain the
Curie constant as 
\begin{equation}
C(x) = C(0) \Big{(}1 + \frac{5x}{3} \Big{)}\, ,
\label{eqC(x)}
\end{equation}
where for $g = 2$
\begin{equation}
C(0) = \frac{3}{4} {\rm \frac{cm^3\,K}{mol}}\ .
\label{eqC(0)}
\end{equation}
Here, a ``mol'' refers to a mole of $\rm{Lu_2V_2O_{7-x}}$
formula units containing two moles of V atoms.  The linear relation with no
adjustable parameters in Eqs.~(\ref{eqC(x)}) and (\ref{eqC(0)}) is plotted as
the straight line in Fig.~\ref{Tc-theta-C}(a), which describes well the Curie
constant data for $x > 0$.  This agreement suggests that increasing the oxygen
vacancy concentration $x$ results in the mixture of localized $S
= 1/2$ and $S = 1$ V spin species dictated by the composition in an ionic
model.  This in turn suggests that the compound does not exhibit metallic
character over our $x$ range.  However, one then expects that the saturation
magnetic moment $\mu_{\rm sat} = g\langle S\rangle \mu_{\rm B} = (1+x)
\mu_{\rm B}$ at $H = 5.5$~T should increase with $x$, instead of remaining
nearly independent of $x$ as seen in Table~\ref{tab-magdata} and
Fig.~\ref{M-sat}.

For a uniform spin $S$ system with the
Heisenberg Hamiltonian
${\cal{H}} =-J\sum_{<ij>}{\vec{S}}_i\cdot{\vec{S}}_j$ with
nearest-neigbor exchange interaction $J$, where the sum is over
nearest-neighbor spin pairs and $J > 0$ corresponds to ferromagnetic
coupling, the mean-field ferromagnetic ordering (Curie) temperature is given
by 
\begin{equation}
T_{\rm C} = \theta = \frac{zJS(S+1)}{3 k_{\rm B}}\, ,
\end{equation}
where $z$ is the nearest-neighbor coordination number of a spin with
neighboring spins.  For a system containing different spin values interacting
with the same $J$ for each nearest-neighbor spin pair, the Curie temperature
at the mean field level is then
\begin{equation}
T_{\rm C} = \frac{zJ\langle S(S+1)\rangle}{3 k_{\rm B}}\, .
\label{eqTc}
\end{equation}
Using Eqs.~(\ref{eqCurieConst2}) and (\ref{eqTc}), one can eliminate $\langle
S(S+1)\rangle$ and express $T_{\rm C}$ in terms of $C$ as 
\begin{equation}
T_{\rm C} = \frac{z J C}{N g^2 \mu_{\rm B}^2}\ .
\label{EqTC}
\end{equation}
Thus one expects that as the Curie constant increases with increasing $x$,
$T_{\rm C}$ should also \emph{increase}.  Remarkably, we find instead from
Table~\ref{tab-magdata} and Fig.~\ref{Tc-theta-C} that both $\theta$ and
$T_{\rm C}$ \emph{decrease} with increasing~$x$.

From Eq.\ (\ref{EqTC}), one can express $J$ in terms of
$T_{\rm C}$ as
\begin{equation}
J = \frac{N g^2\mu_{\rm B}^2 T_{\rm C}}{zC}\ .
\label{eqJ}
\end{equation}
Setting $N$ to be twice Avogadro's number (there are two V atoms per formula
unit in $\rm{Lu_2V_2O_{7-x}}$), $g = 2$, $z = 6$ from the structure, and
taking $C$ from Eqs.\ (\ref{eqC(x)}) and (\ref{eqC(0)}), Eq.\ (\ref{eqJ})
gives 
\begin{equation}
J = 0.500\ \frac{T_{\rm C}}{C}\, ,
\label{eqJ2}
\end{equation}
where $J$ and $T_{\rm C}$ are in units of K and $C$ is in units of
cm$^3$\,K/mol.  Using Eq.\ (\ref{eqJ2}) and the $T_{\rm C}(x)$ and $C(x)$
data in Table~\ref{tab-magdata}, $J(x)$ was computed and is plotted in
Fig.~\ref{Tc-theta-C}(b).  One sees that $J$ monotonically and strongly
decreases with increasing $x$, by a remarkable factor of about 6
as $x$ increases from 0 to 0.65\@.

\section{Summary and Conclusions\label{Summary}}

We have synthesized and studied the new defect pyrochlore series
${\rm Lu_2V_2O}_{7-x}$ with $x$ = 0.40--0.65 carried out in an attempt to dope
the parent ($x=0$) ferromagnetic semiconductor $\rm{Lu_2V_2O_{7}}$
into the metallic state.   The compound $\rm{Lu_2V_2O_{7}}$ contains two
crystallographically inequivalent oxygen sites denoted as O and O$^\prime$
sites, and the composition can be written as $\rm{Lu_2V_2O_{6}O^\prime}$.  For
$x \gtrsim 0.5$, significant Lu-V antisite disorder clearly occurs that
increases roughly as $\sim x^2$.  This is consistent with our structural
model for the oxygen vacancies, where oxygen depletion occurs at the
O$^\prime$ site which is tetrahedrally coordinated by Lu in stoichiometric
$\rm{Lu_2V_2O_{6}O'}$.  This model also appears to explain the anomalous
nonmonotonic dependence of the lattice parameter on $x$ which initially
increases with $x$ for $0 < x
\lesssim 0.4$, and then decreases for $x \gtrsim  0.4$.  We suggest that the
Lu$_4$O$^\prime$ tetrahedra (and the unit cell) first expand as O$^\prime$
atoms are removed from the centers of the Lu$_4$ tetrahedra, via elimination
of the Lu-O$^\prime$-Lu bonding within the affected Lu$_4$ tetrahedra, and
then the unit cell subsequently shrinks with further increase in $x$ due to
the steadily reduced average oxygen coordination number of the Lu$^{+3}$
cations.  Our structural model for the oxygen vacancy position should be
testable via future neutron diffraction studies of the O and O$^\prime$ site
occupancies versus $x$ in Lu$_{2}$V$_{2}$O$_{7-x}$ when samples of sufficient
mass are synthesized.

Recent studies on Nb substitution effects, single crystal
polarized neutron scattering measurements, and $^{51}$V NMR studies
 indicate that a ferro-orbital
ordered state accounts for the simultaneous ferromagnetic and
semiconducting behaviour of the undoped ${\rm Lu_2V_2O_{7}}$ parent
compound.\cite{Shamoto2002,Ichikawa2005,Kiyama2006}  The oxygen defect
pyrochlore series  ${\rm Lu_2V_2O}_{7-x}$ ($x = 0.40$--0.65) studied here
remains ferromagnetic throughout.  Unfortunately, the powder nature of the
samples prevented us from carrying out conventional electronic transport
measurements to determine whether the compounds are insulating or metallic
(as $T\to 0$). The observed Curie constant $C$ in the high-temperature
Curie-Weiss behavior of the magnetic susceptibility \emph{increases} rapidly
with increasing $x$, approximately following the mean field prediction for the
mixture of localized $S = 1/2$ (${\rm V^{+4}}$) and $S = 1$ (${\rm V^{+3}}$)
spins dictated by an ionic model for the composition $x$, suggesting the
absence of metallic character of the material over our $x$ range.  However,
the ferromagnetic ordering temperature $T_{\rm C}$ and
Weiss temperature $\theta$ both strongly \emph{decrease} monotonically with
increasing~$x$, which is opposite to the behaviors predicted from
$C(x)$.  Furthermore, the high-field (5.5~T) saturation moment at low
temperatures (5 K) is nearly independent of
$x$ [$\mu_{\rm sat}
\approx 1\ \mu_{\rm B}$/(V~atom), as expected for $S = 1/2$] and does not
show the expected increase with increasing $x$, $C$ and $\langle S\rangle$. 
These latter three anomalous behaviors suggest that the hole-doped ${\rm
Lu_2V_2O}_{7-x}$ series may actually be metallic, contrary to expectation
from the observed $C(x)$, so that both localized ($S = 1/2)$ and itinerant
\emph{d}-electrons may coexist.  In that case one would
need to explain why the conduction carriers give an apparent Curie-Weiss
contribution to the magnetic susceptibitity (in addition to that from the
localized V$^{+4}$~spins 1/2) that increases with increasing $x$.  These
issues will be very interesting to examine further in future experimental and
theoretical studies.

\begin{acknowledgments}

Ames Laboratory is operated for the U.S. Department of Energy by Iowa State
University under Contract No.\ W-7405-Eng-82. This work was supported by the
Director for Energy Research, Office of Basic Energy Sciences. 
\end{acknowledgments}


\begin{thebibliography}{99}
\bibitem{misnomer}The A or B sublattice of the pyrochlore structure is itself
often called ``the pyrochlore structure'' by physicists, a misnomer.  The
identical sublattice structure also occurs in the B sublattice of the
cubic A[B$_2$]X$_4$ spinel structure and in the B sublattice of the AB$_2$
C-15 cubic Laves phase structure.
\end{thebibliography}
\end{document}